\documentstyle[12pt,aaspp4]{article}

\lefthead{Reid et al}
\righthead{Faint M-dwarfs}
\begin {document}
\title{ Faint M-dwarfs and the structure of the Galactic disk
\footnote
{Based partly on observations obtained at the W.M. Keck Observatory, which
is operated jointly by the California Institute of Technology and the
University of California, partly on data from the 200-inch telescope
at Palomar mountain, owned and operated by the California Institute
of Technology, and partly on data from 
the 60-inch telescope at Palomar
Mountain which is jointly owned by the California Institute of Technology
and the Carnegie Institution of Washington}}

\author {I. N. Reid, J. E. Gizis, J. G. Cohen, M. A. Pahre}
\affil {Palomar Observatory, 105-24, California Institute of Technology, Pasadena, CA 91125,
e-mail: inr@astro.caltech.edu, jeg@astro.caltech.edu, jlc@astro.caltech.edu, map@astro.caltech.edu}
\author {D. W. Hogg}
\affil {Theoretical Astrophysics, 130-33, California Institute of Technology, Pasadena, CA 91125,
e-mail: hogg@tapir.caltech.edu}
\author {L. Cowie, E. Hu, A. Songaila}
\affil {Institute for Astronomy, University of Hawaii, 2680 Woodlawn Drive, Honolulu, HI 96822,
e-mail: cowie@ifa.hawaii.edu, acowie@ifa.hawaii.edu, hu@ifa.hawaii.edu}
\begin {abstract}

We use broadband photometry and low-resolution spectra of
a complete sample of late-K and M dwarfs brighter than I=22 in three
fields at high galactic latitude to study issues relating to galactic 
structure and large scale abundance gradients in the Galaxy.
The observed starcounts in each field are a good match to 
the predictions of models based on deep starcount data in other
intermediate-latitude fields, 
and these models identify the late-type stars as members of the
Galactic disk.
Abundances for these late type stars are estimated via
narrowband indices
that measure the strength of the TiO and CaH bands in their spectra.
Our results show that the average abundance
in the Galactic disk remains close to solar even at heights of
more than 2 kpc above the Plane.

\end {abstract}

\keywords { stars: late type, Galaxy: Abundances, Galaxy: Structure }

\section {Introduction }

It is by now generally agreed that the vertical distribution of stars in the
Galactic disk cannot be represented by a simple, single-component density
law. Deep, wide-field optical number-magnitude starcounts show that the
slope of the overall density distribution flattens by a factor of $\sim4$
between 1.2 and 1.5 kiloparsecs above the Plane, well before the halo
becomes the dominant population at $z \sim 5$ kpc (Gilmore \& Reid, 1983;
Majewski, 1992; Reid \& Majewski, 1993 - hereinafter, RM93; Gould, Flynn
\& Bahcall, 1996). What is not clear, however, is how one interprets
these `extra' stars, both in terms of deconvolving the observed
distribution into thin and thicker components (Majewski, 1993; 
Robin, 1994) and in deciding the physical significance of the deconvolved
components. 

Majewski (1993) outlines eight different scenarios for the
formation of what we will refer to as the intermediate population (or IP II),
and one can characterise the extremes as being those models which treat
the IP II as the result of a separate, distinct star formation event
(e.g. Wyse \& Gilmore, 1988; Robin et al, 1996) and those which identify 
the IP II as the high-velocity tail of the old disk (Norris, 1987).
One method of constraining these various hypotheses is through measurement of the
abundance distribution of the stars in the IP II, in particular determining
whether there is a significant gradient in the mean abundance with height
above the Galactic Plane. To date, however, the requisite observational data 
are still relatively sparse. Yoss, Neese \& Hartkopf (1987) use observations of
high-latitude K-giants to derive a metallicity gradient of $\sim -0.4$ dex kpc$^{-1}$
to z=700 pc. and $\sim -0.18$ dex kpc$^{-1}$ thereafter (to 8 kpc), 
although their sample includes relatively few stars above 2.2 kpc and 
shows substantial dispersion. Trefzger, Pel and Gabi (1995) and
Gilmore, Wyse \& Jones (1995) have surveyed F and G dwarfs towards the
South Galactic Pole, using VBLUW photometry and medium resolution,
low signal-to-noise spectra respectively. Again, neither study include
many stars above 2.5 kiloparsecs. Trefzger et al find a relatively
steep gradient close to the Plane ($\sim -0.55$ dex kpc$^{-1}$), and
both match the K giant survey in finding little variation in the mean 
abundance above $\sim 1$ kiloparsec, although Gilmore et al's data
show significantly more dispersion. Finally, Majewski's (1992) UBV
data extend to z $>$ 10 kpc and suggest a somewhat higher mean abundance
below 6 kiloparsecs.

In this paper we present the first attempt to use observations of
distant late-type K and M dwarfs to study this question. At least 80 \% of
the stars in the Galactic disk are M dwarfs, so these stars have a high
surface density at faint magnitudes.  Moreover M dwarfs can be recognized
even with low dispersion spectra.  At the faint magnitude level under
consideration, any M star seen must be an M dwarf; high luminosity M
giants and supergiants would be so far out in the galactic halo that
even if they were to be found there in the same ratio as K giants
and K supergiants (which is unlikely given the higher metallicity required 
to produce M giants), their surface density would be insignificant compared to the number
of bona fide M dwarfs.

Furthermore low-abundance M-type
subdwarfs can be identified through the enhanced strength of the hydride
bands (particularly CaH) relative to the TiO absorption (Bessell, 1982).
Until recently, the complex, diatomic-molecule dominated spectra of these
stars defied quantitative abundance analysis, but Gizis (1997) has shown that
the atmosphere models developed by Allard \& Hauschildt (1995) can be
combined with narrowband spectrophotometry to derive an approximate ($\sim\pm
0.3$ dex) abundance scale for these stars. Given these results, we have 
undertaken a preliminary study, analysing Keck LRIS data of a magnitude-limited 
sample of faint M-dwarfs. Section 2 describes the observational data; section 3
compares the observed colour-magnitude diagrams against Galactic starcount
models and discusses the abundance distribution; section 4 summarises our 
conclusions.

\section{Observations}

Our  sample consists of the K and M dwarfs drawn from three fields:
the 0-hour deep field (l=125$^o$, b=-50$^o$); SSA 13 (l=109$^o$, b=74$^o$);
and SSA 22 (l=63$^o$, b=-44$^o$). The first field is $7.3 \times 2$ arcminutes
in size (0.004 square degrees), while the other two fields are $6' \times 2'$
(0.003 sq. deg.). All three fields are part of a deep, K-band limited
redshift survey being undertaken by groups at Caltech (Cohen et al, 1996)
and Hawaii (Cowie et al, 1996).   The primary goal is a deep redshift
survey of extragalactic objects, but objects within the magnitude limits
were observed irrespective  of morphology.
Here we concern ourselves with the
galactic stars found during the course of these redshift surveys.
This paper discusses observations of the complete sample of 68
stars with $I<22$ mag. in these fields.

The spectroscopic sample consists of those objects with observed spectra
judged to have sufficient signal-to-noise ratios to be useful in
deducing abundances for these K and M dwarfs.  It includes 43 stars.
The spectroscopic observations of the 0 hour field sources are described 
in detail by Cohen et al (1997), while Cowie et al (1996) discuss the data
for objects in the two Hawaii fields.

A detailed description of the observations and analysis of the $U, B, V, R, I$,
and $K$ photometry for the 14 stars in the 0 hour field can be found in
Pahre et al (1997).   The same can be found for the Hawaii fields
(observed only at $B, I$, and $K$) in Cowie et al (1996). We have supplemented
the SSA22 data with BVI photometry obtained with LRIS on the Keck telescope
in September, 1996, calibrating the observations using the same methods outlined
by Pahre et al (1997). These observations included shorter-exposure images,
and hence avoided saturating the brighter stars. We have used the IRAS 100 $\mu m$ maps
to estimate the interstellar absorption in each field. Using the calibration
given by Laureijs, Helou \& Clark (1994), we estimate a total extinction of
A$_V \sim 0.22$ mag in the 0-hour field, corresponding to A$_I \sim 0.10$, 
E$_{B-I} \sim 0.19$ and E$_{I-K} \sim 0.08$ mag; A$_V \sim 0.23$ mag in SSA22;
and negligible extinction (A$_V \sim 0.005$ mag) in SSA13. We assume
that all of the extinction lies in the immediate vicinity (within 200 pc.)
of the Sun. Figure 1 shows the reddening-corrected two-colour ($B-I$)/($I-K$) 
diagram for all of the spectroscopically-confirmed stars in these fields.

All of the objects were observed spectroscopically using multislit
masks with LRIS (Oke et al, 1994) on the Keck 10-metre telescope.
The slits used were 1.4 arcseconds wide , and a 300 l/mm
grating gives spectra of 16 \AA ($\sim 5$ pixels) resolution covering
the wavelength range $\sim 5000$ to 10000 $\AA$. (The actual coverage
on each spectrum depends on the exact position of the slit within the
rectangular field.) Full details of the reduction procedures are given
in Cowie et al (1996) and in Cohen et al (1997).  The spectra from the
0 hour field were set on a flux scale using observations of the white dwarf standard 
G191-B2B (Oke 1990).
This star was observed using a 1.5 arc-sec aperture longslit in LRIS.  While the calibration
takes out the overall spectral response of the CCD, varying losses of light at the slit
lead to the calibrations for individual multi-slits being
uncertain by up to 10 \%  over broad wavelength intervals of $\sim 2000 \AA$
or more. However, the widest of our narrowband indices spans only 250 \AA, so
the uncertainties in the indices are dominated by the signal-to-noise in the individual 
spectra rather than by systematic errors in the flux calibration.

\section{Discussion}

\subsection{Starcounts}

We have compared the observed colour-magnitude distribution of the photometric
sample of sixty-eight stars with I$<$22 mag. against the predictions of
starcount models. As the starting point for our analysis, we take model B
from Reid {\sl et al} (1996 - hereinafter RYMTS). This model includes 
three stellar components: the old
disk, modelled as a sech$^2 {h \over h_0}$ distribution, with $h_0$ of 350 parsecs;
an intermediate population, with a local normalisation of 5 \% relative to the
old disk and a sech$^2$ density law, $h_0 = 1500$ pc.; and an r$^{-3.5}$ halo with
the local density normalised to 0.15 \% that of the old disk. We have also adopted
a scalelength of 2.5 kiloparsecs for the radial density law in the Galactic disk
(Robin, Cr\'ez\'e \& Mohan, 1992). The colour-magnitude relations adopted for
these three populations are from photometry of nearby stars, of members of the open
cluster NGC 2420 and of stars in the globular cluster NGC 6752, respectively.
Full details are
given in RYMTS, who show that this model reproduces both 
the colour-magnitude distribution in the intermediate-latitude PSR 1640 field
(l=41$^o$, b=38$^o$) and the density distribution towards the South Galactic Pole. 

Figure 2a compares the overall starcount predictions of model B against the
observations in the three fields considered in this paper. The agreement is
well within the formal Poissonian uncertainties.
Limiting the sample to the magnitude range 17 $< I < 22$, we
observe 19, 11 and 37 stars in the 0 hour field, SSA13 and SSA22.
The predicted numbercounts are 15, 11 and 32 stars in the respective fields. 

Figure 2b compares the colour-magnitude distributions predicted by this model
against the observed (I, (I-K)) distribution in each field. For greater clarity,
the predicted starcounts in each field are matched to a solid angle 
of 0.01 square degrees - a factor of 2.5 greater than the 0-hour field and 
$\sim3$ greater than the SSA 13 and SSA 22. The model predicts two distinct
sequences: late F- and G-type halo subdwarfs ( (I-K) $<$ 1.5) lying at a distance 
of $\sim$ 10 kiloparsecs; and a sequence of late-K and M-type disk dwarfs, mainly
contributed by the IP II, 
at distances of from $\sim 1$ to 4 kiloparsecs. The small scaleheight of the old
disk component leads to a low surface density and an insigificant contribution to these 
faint starcounts, even in the intermediate-latitude field SSA22. 

As described in RYMTS, the convolution of the density law and the sampling
volume lead to a preferred distance modulus for a given stellar populaton. Since
the old disk has the steepest vertical density law and the halo the shallowest,
the former stars are sampled primarily at (m-M)$\sim$ 7 magnitudes, while the
latter lie preferentially at (m-M)$\sim$ 15 magnitudes. Thus, at a given 
apparent magnitude, the halo contributes more luminous (bluer) stars than
the disk. The models plotted in figure 2b clearly show that the later-type
stars, which contribute the bulk of the stars in each field, are
members of the Galactic disk.

\subsection{The abundance distribution}

The primary aim of this paper is to use the late-type dwarfs in our sample to probe
the abundance distribution at large distances above the Plane. In order to
do this we need to estimate both abundances and distances. We have used B-,
I- and K-band data for individual stars within 8 parsecs to define 
(M$_I$, (B-I)) and (M$_I$, (I-K)) relations (figure 3). The curves fitted are sixth order
polynomials, and the dispersion is $\sigma_I = 0.35$ magnitudes in the (B-I)
calibration and $\sigma_I = 0.52$ magnitudes for (I-K). Comparing the distance
estimates for individual stars, the mean difference between the (B-I) and
(I-K) photometric parallaxes is 40 \% of the distance based on
averaging the two estimates, with
no evidence of a systematic offset between the two distance scales. We
have therefore taken the straight average of the two estimates as the appropriate distance
for each star and adopt an uncertainty of $\sim 30 \%$. This is 
adequate for the present purposes.

Gizis (1997) has
shown that abundance estimates for late-K and M-type dwarfs can be derived
from the relative strengths of the calcium hydride band at $\sim 6800 \AA$
and the titanium oxide $\gamma (0,0)$ band at $\sim 7050 \AA$. Reid, Hawley
\& Gizis (1995 - RHG) have defined several narrowband indices designed to measure
the strengths of these bands, and figure 4 plots data for two of the CaH
indices against TiO5, the full depth of the TiO bandhead. In addition to
data (from RHG) for single disk stars within 8 parsecs of the Sun, we have plotted
Gizis' measurements of the band-strength in intermediate and extreme subdwarfs.
A comparison with the Allard and Hauschildt models shows that the former stars, 
plotted as solid square in figure 4, have an abundance of $[Fe/H] \sim -1$
while the latter are significantly more metal-poor, with $[Fe/H] \sim -2$. 
We have taken nearby K and M dwarfs to be representative of a solar-abundance
population.

We have fitted second-order polynomials to each of the three sets of stars
in figure 4 and used these as a grid to match against narrowband indices
derived from our observations of the faint programme stars. Table 1 lists the
band-strength data for the later-type stars in the current sample.
Broadband photometry and positions for stars in SSA 13 and SSA 22 
are given by Cowie et al (1996), while Pahre et al (1997) present
UBVRIK for stars in the 0-hour field. Figure 5 plots
the resulting bandstrengths and it is clear that the IP II stars in our sample lie
closest to the polynomial fit to the local (solar abundance) stars, consistent
with previous observations of IP II F and G dwarfs. We have attempted to
set our abundance estimates onto a more quantitative scale by using the
[Fe/H]=0 (nearby stars) and [Fe/H]=$-$1 (intermediate subdwarfs) sequences
as fiducial references. Defining CaH$^0$ and CaH$^{-1}$ as the values 
predicted by the solar and [Fe/H]=$-$1 relations for a given bandstrength, TiO5,
we calculate the quantity

$$ \Delta_C \qquad = \qquad { {CaH^{obs} - CaH^0} \over {CaH^0 - CaH^{-1}}} $$

\noindent
where CaH$^{obs}$ is the observed strength of the CaH feature for the star
in question.
This scaling compensates for the changing separations of the two
fiducial sequences, although without an adequate number of calibrators
with high-precision abundance measurements, we can only assume that the
variation in $\Delta_C$ with [Fe/H] is similar throughout the full
temperature range covered here.

Using the distances derived from the average of the photometric parallaxes,
we have computed the
distance, z, above the Galactic Plane for each star. Limiting the
sample to stars with significant TiO absorption (TiO5 $< 0.85$), figure 6 plots
the scaled CaH indices against z. The average values of the two indices
if we include all stars are $\Delta_{c2}$ = $-$0.16 and $\Delta_{c3}$ = $-$0.06 
(24 stars), while we derive $\Delta_{c2}$ = $-$0.27 and $\Delta_{c3}$ = $-$0.06
for the 15 stars with $z < 2000$ parsecs. As yet, we do not have
sufficient calibrating observations of stars of known abundance to be able
to transform these measured offsets into real abundances, but the present results
are clearly suggestive of values closer to those of the old disk than
previously suggested.

\section{Conclusions}

We have analysed photometry and low-resolution spectroscopy of faint
late-K and early M-dwarfs in three intermediate- and high-latitude
fields. A comparison with the predictions of a starcount model
which matches the faint (R$< 24$) stellar colour-magnitude distribution 
in two other fields (Reid et al, 1996) shows that
these stars, lying at distances of up to 2 kiloparsecs above the Plane 
and members of the intermediate population. Our analysis, based
on the strength of CaH and TiO molecular bands, demonstrates that these stars
have abundances close to that of the old disk. Clearly, 
these results are preliminary, both given the total number of stars
in our sample and the approximate nature of the calibration of the measured 
bandstrength against real abundances. However, our analysis does 
indicate a possible new direction that can be taken in  exploring the abundance distribution
within Galactic stellar populations.

\acknowledgments 

The Keck telescope project was made possible by a generous grant from the W.M. Keck
Foundation. INR's contribution to this research was supported partially by NSF 
grant AST-9412463; MAP was supported by NSF award AST-9157412 and the Bressler Foundation;
DWH was supported by NSF award AST-9529170.

\clearpage



\makeatletter
\def\jnl@aj{AJ}
\ifx\revtex@jnl\jnl@aj\fi
\makeatother


\begin{deluxetable}{rrrrrrrr}
\tablecaption{Spectroscopic indices \label{table1}}

\tablewidth{0pt}
\tablenum{1}
\tablehead{
\colhead{Star} & 
\colhead{TiO5} &
\colhead{CaH2} &
\colhead{CaH3} }
\startdata
 & 0 hour field &\nl
PD0K 03 & 0.765 & 0.604 & 0.816 &\nl
PD0K 04 & 0.792 & 0.627 & 0.837 &\nl
PD0K 09 & 0.550 & 0.494 & 0.763 &\nl
PD0K 16 & 0.643 & 0.503 & 0.790 &\nl
PD0K 19 & 0.597 & 0.597 & 0.811 &\nl
PD0K 21 & 0.458 & 0.428 & 0.726 &\nl
PD0K 22 & 0.549 & 0.444 & 0.712 &\nl
PD0K 49 & 0.387 & 0.419 & 0.693 &\nl
PD0K 50 & 0.625 & 0.626 & 1.003 &\nl
PD0K 91 & 0.549 & 0.479 & 0.718 &\nl
PD0K 97 & 0.333& 0.348 & 0.594 &\nl
PD0K 101 & 0.392 & 0.597 & 0.689 &\nl
 & \nl
 & SSA 13 &\nl
17 & 0.461 & 0.433 & 0.697 &\nl
35 & 0.435 & 0.347 & 0.655 &\nl
73 & 0.417 & 0.508 & 0.808 &\nl
84 & 0.378 & 0.321 & 0.583 &\nl
97 & 0.583 & 0.466 & 0.709 &\nl
 & \nl
 & SSA 22 &\nl
24 & 0.290 & 0.333 & 0.387 &\nl
61 & 0.644 & 0.400 & 0.845 &\nl
112 & 0.400 & 0.475 & 0.754 &\nl
157 & 0.489 & 0.477 & 0.755 &\nl
169 & 0.615 & 0.435 & 0.713 &\nl
171 & 0.601 & 0.436 & 0.706 &\nl
200 & 0.616 & 0.480 & 0.779 &\nl
\enddata
\end{deluxetable}


\clearpage

\begin{figure}
\plotfiddle{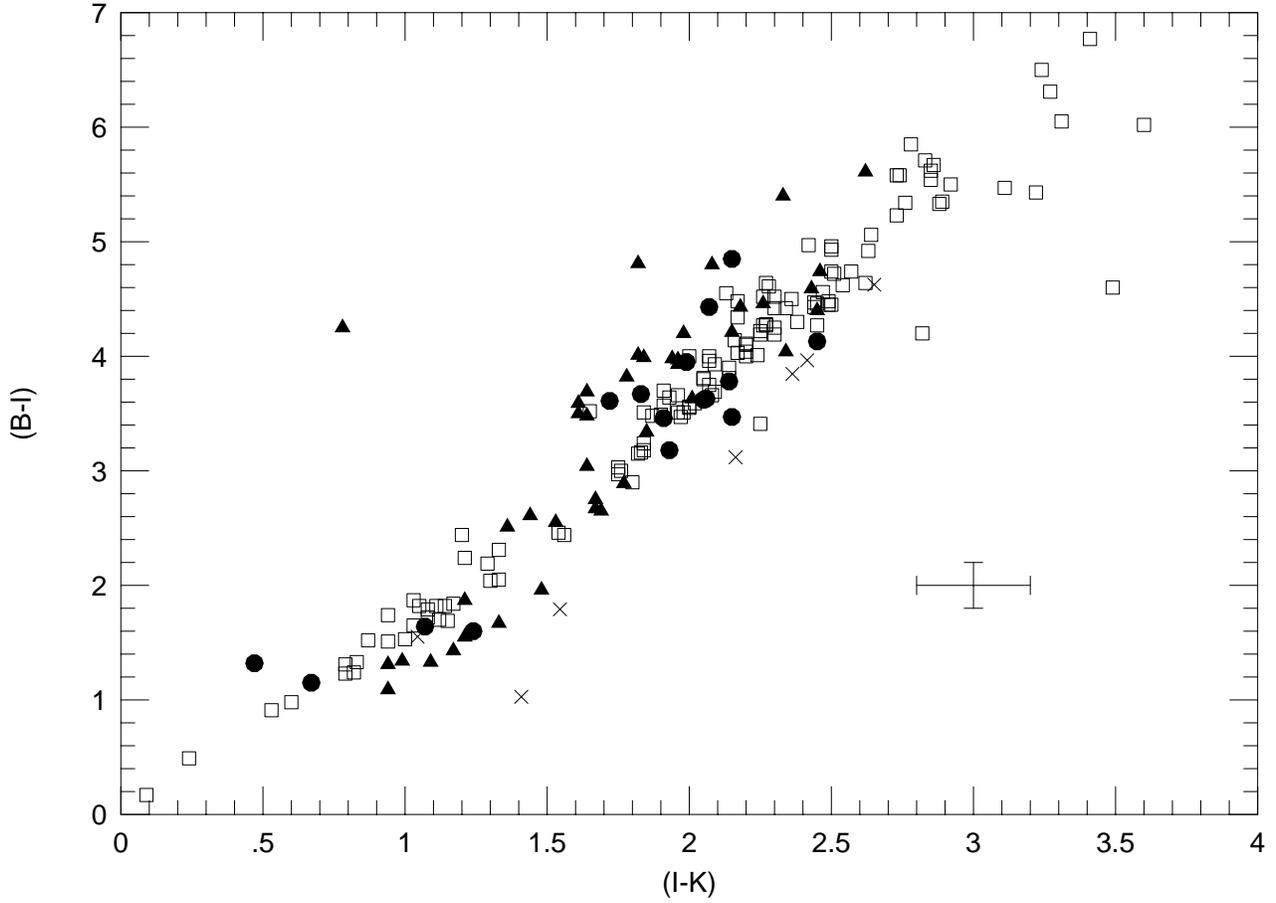}{7truein}{0}{70}{70}{-275}{50}
\caption{
The (B-I)/(I-K) two-colour diagram for stars in the 0-hour field
(solid dots), SSA 13 (crosses) and SSA 22 (solid triangles).
The reference sequence is provided by Leggett's (1992) photometry of stars
within 8-parsecs of the Sun (plotted as open squares). The error-bar shows 
representative uncertainties for I$\sim$ 21 magnitude and fainter.
\label{fig1}
}
\end{figure}

\begin{figure}
\plotfiddle{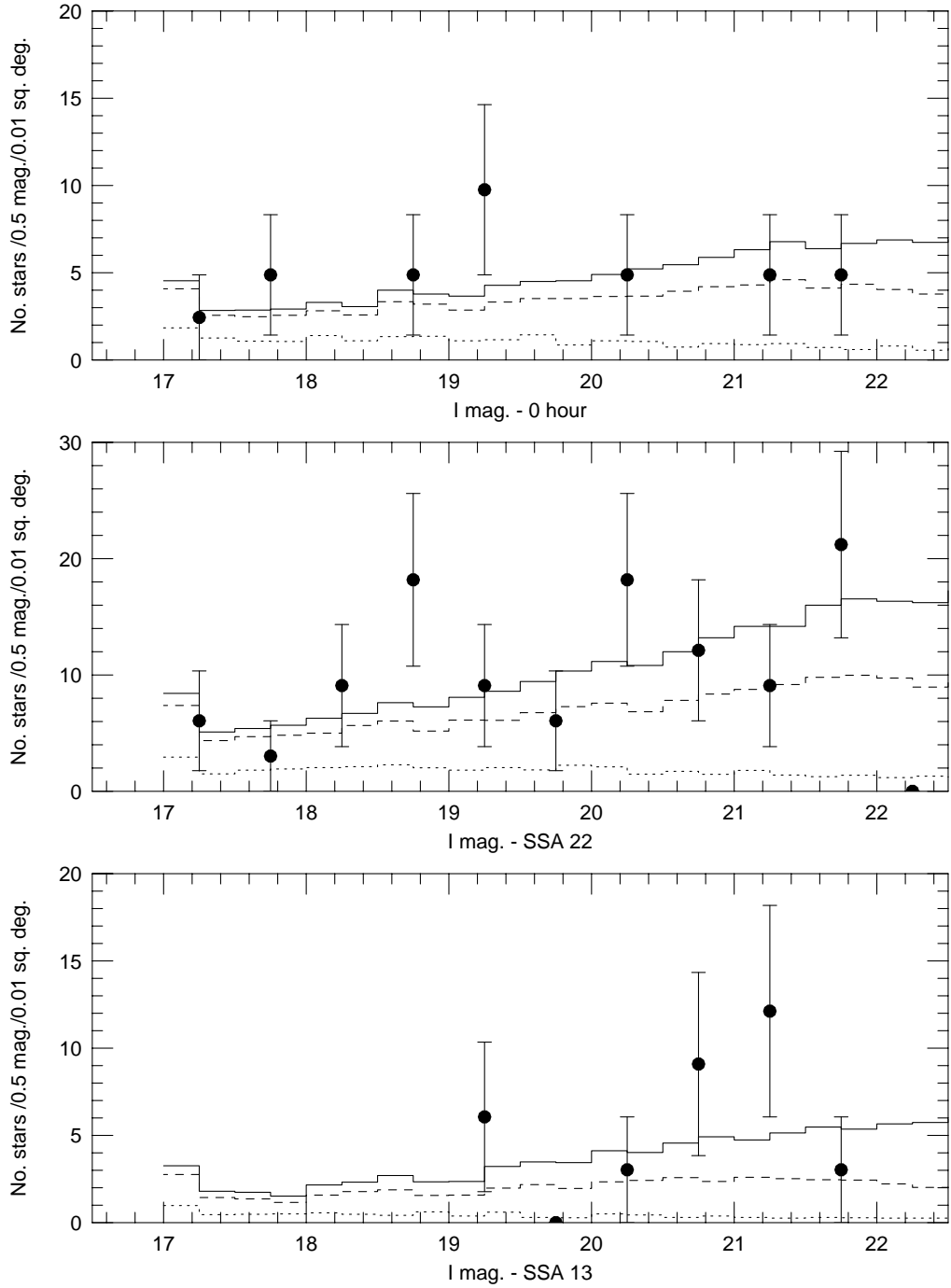}{7truein}{0}{80}{80}{-250}{-50}
\caption{
(a) Predicted and observed number counts for the three fields
in this study. The solid histogram outlines the total counts; the
dotted line marks the contribution of the old disk; and the dashed
line separates the contribution from the IP II and halo.
\label{fig2a}
}
\end{figure}

\setcounter{figure}{1}

\begin{figure}
\plotfiddle{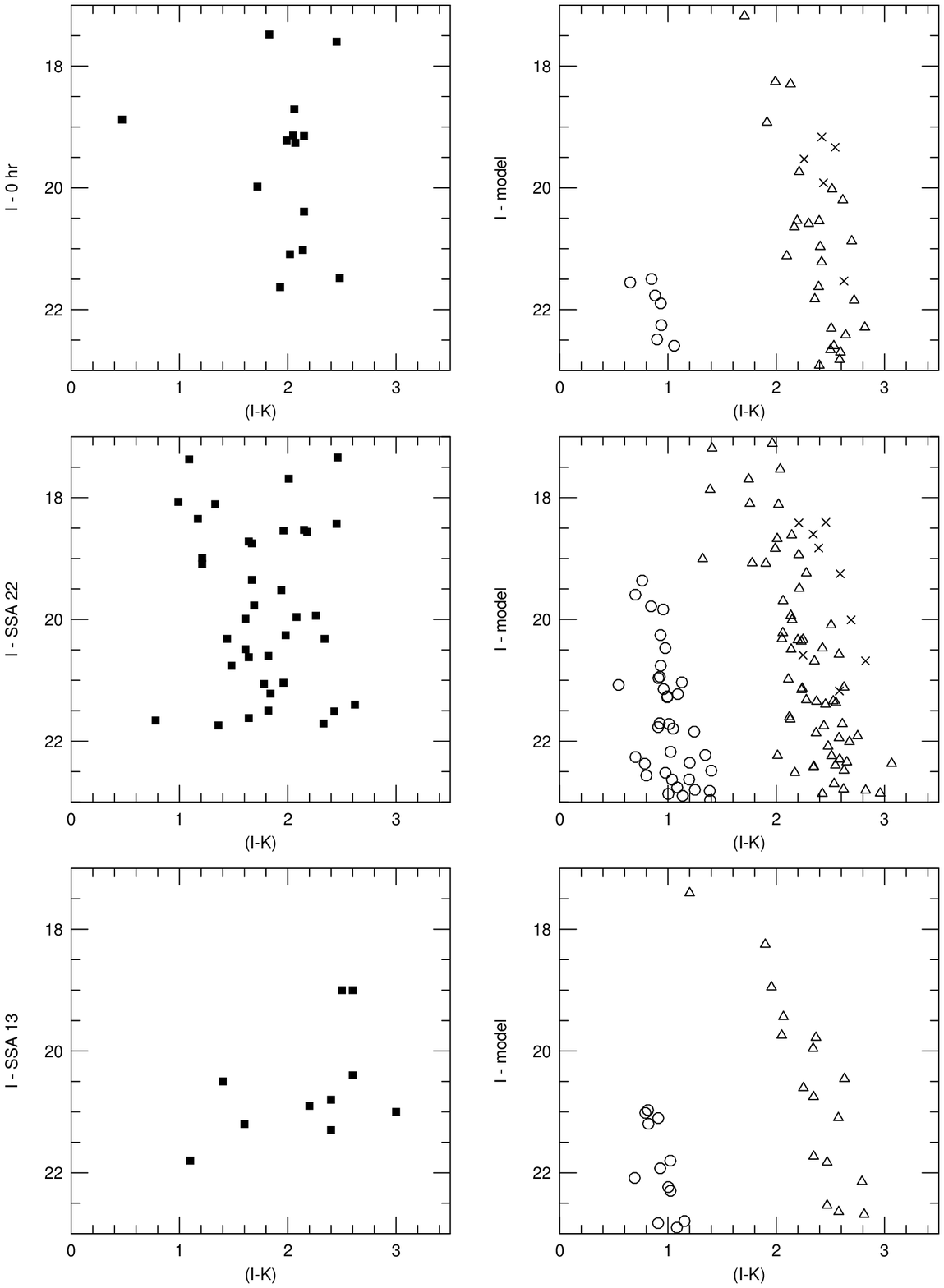}{7truein}{0}{80}{80}{-250}{-50}
\caption{
(b) Predicted and observed (I, I-K) colour magnitude diagrams. 
The right-hand panel show the predictions
for each field (and for a solid angle of 0.01 square degrees) made by 
the model described in the text, where the red sequence is due to old disk (crosses)
and IP II stars (triangles) and
the blue due to halo stars(open circles). 
The left-hand panels plot the observed distributions.
\label{fig2b}
}
\end{figure}

\begin{figure}
\plotfiddle{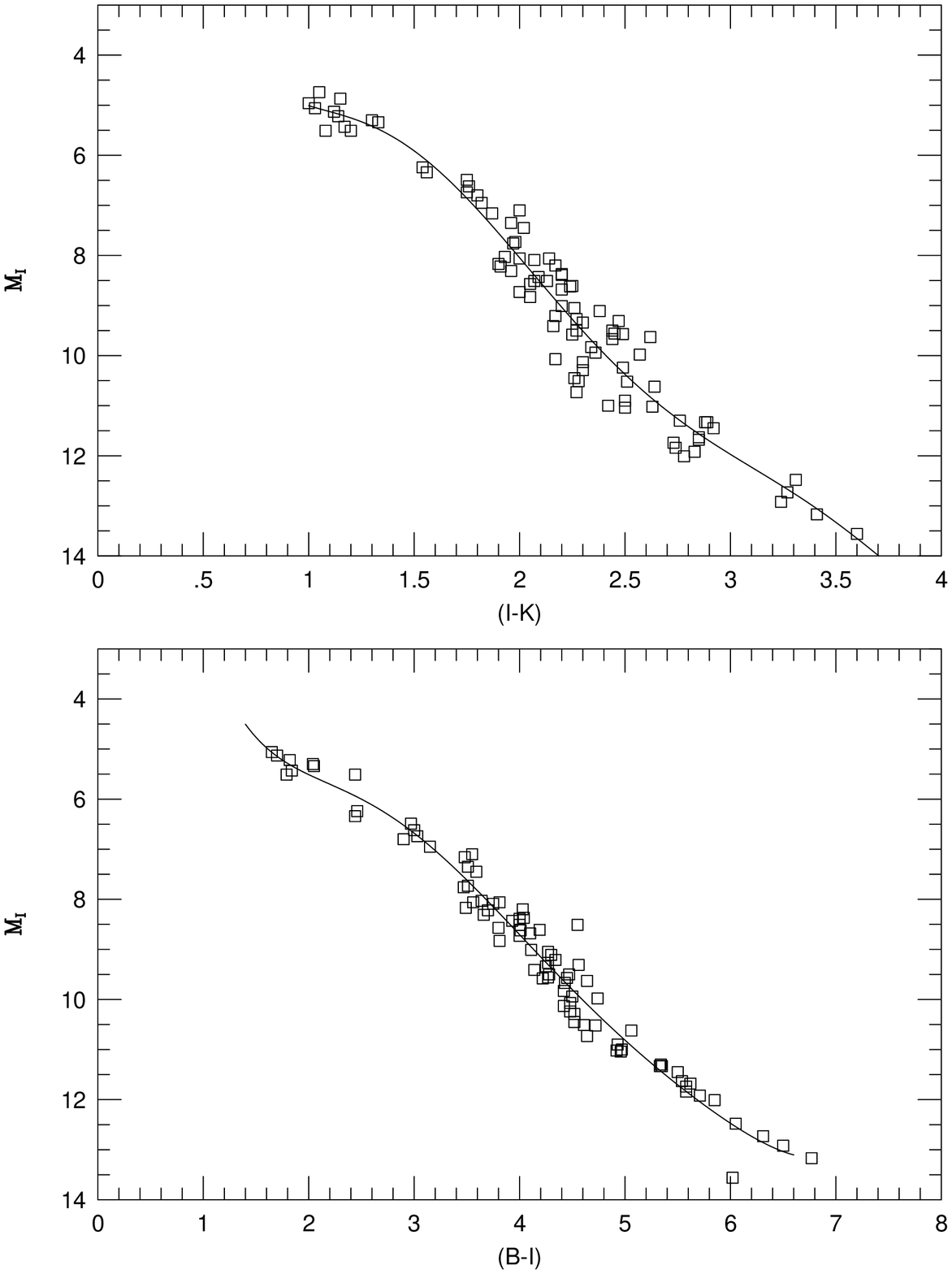}{7truein}{0}{80}{80}{-250}{-50}
\caption{
The (M$_I$, (I-K)) and (M$_I$, (B-I)) colour-magnitude diagrams 
defined by stars within 8-parsecs of the Sun. The majority of the photometry 
is from Leggett (1992). In each case the line shows mean relation defined by
a sixth order polynomial fitted to the data.
\label{fig3}
}
\end{figure}

\begin{figure}
\plotfiddle{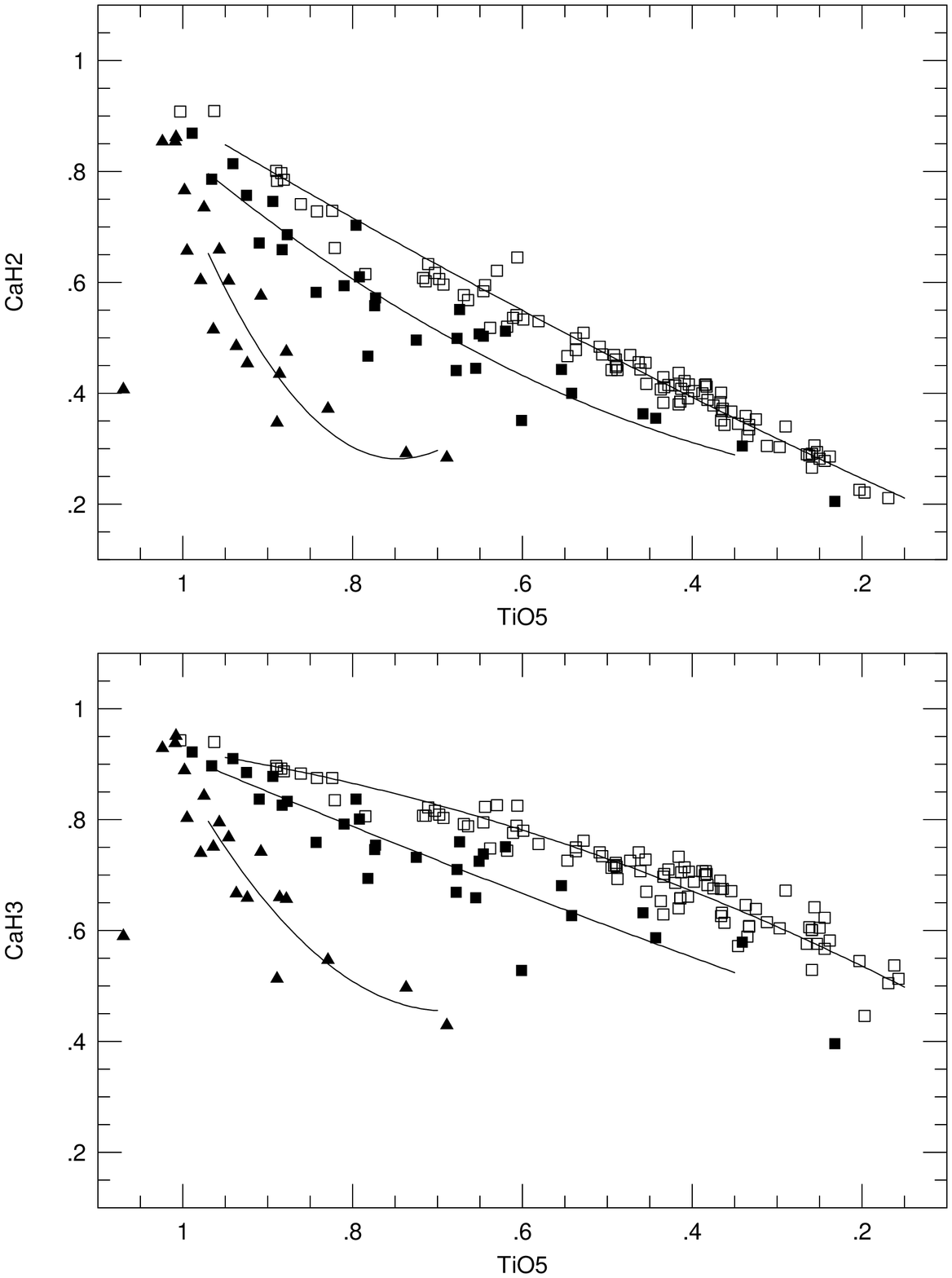}{7truein}{0}{80}{80}{-250}{-50}
\caption{
Calcium hydride strength as a function of the TiO $\gamma (0, 0)$
bandstrength. Data for single stars within 8 parsecs of the Sun (from RHG)
are plotted as open squares. These define a solar-abundance sequence. The
solid symbols are subdwarfs from Gizis (1996), with the squares being
`normal' subdwarfs with [Fe/H] $\sim -1$ and the triangles `extreme'
subdwarfs, [Fe/H] $\sim -2$. Second-order polynomials fitted to each
sequence are also shown.
\label{fig4}
}
\end{figure}

\begin{figure}
\plotfiddle{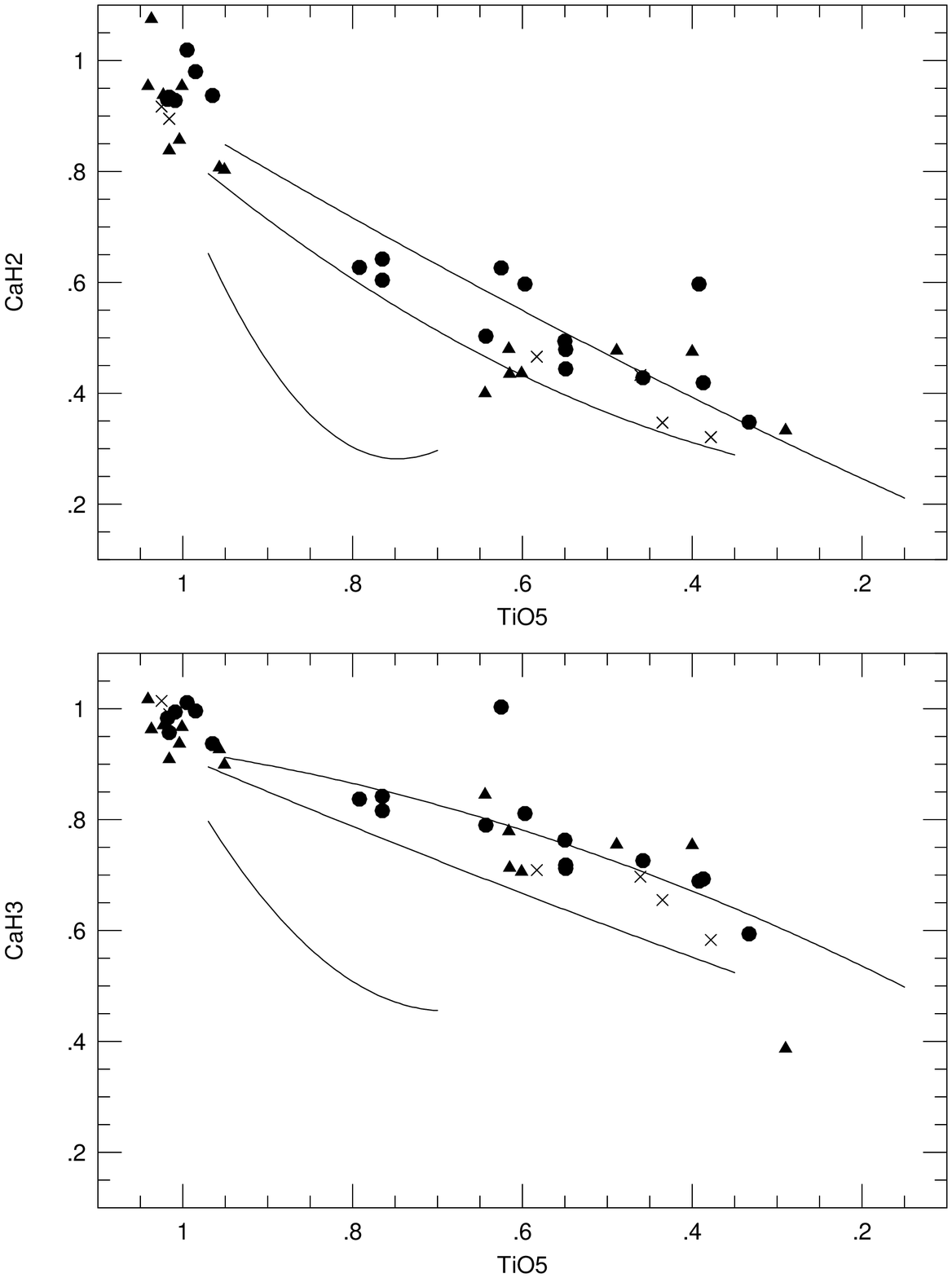}{7truein}{0}{80}{80}{-250}{-50}
\caption{
Calcium hydride and titanium oxide bandstrengths for the
programme stars in our sample. The data are coded as in figure 1 and
the fiducial lines are the polynomial fits to the disk dwarfs, intermediate
subdwarfs and extreme subdwarfs from figure 4.
\label{fig5}
}
\end{figure}

\begin{figure}
\plotfiddle{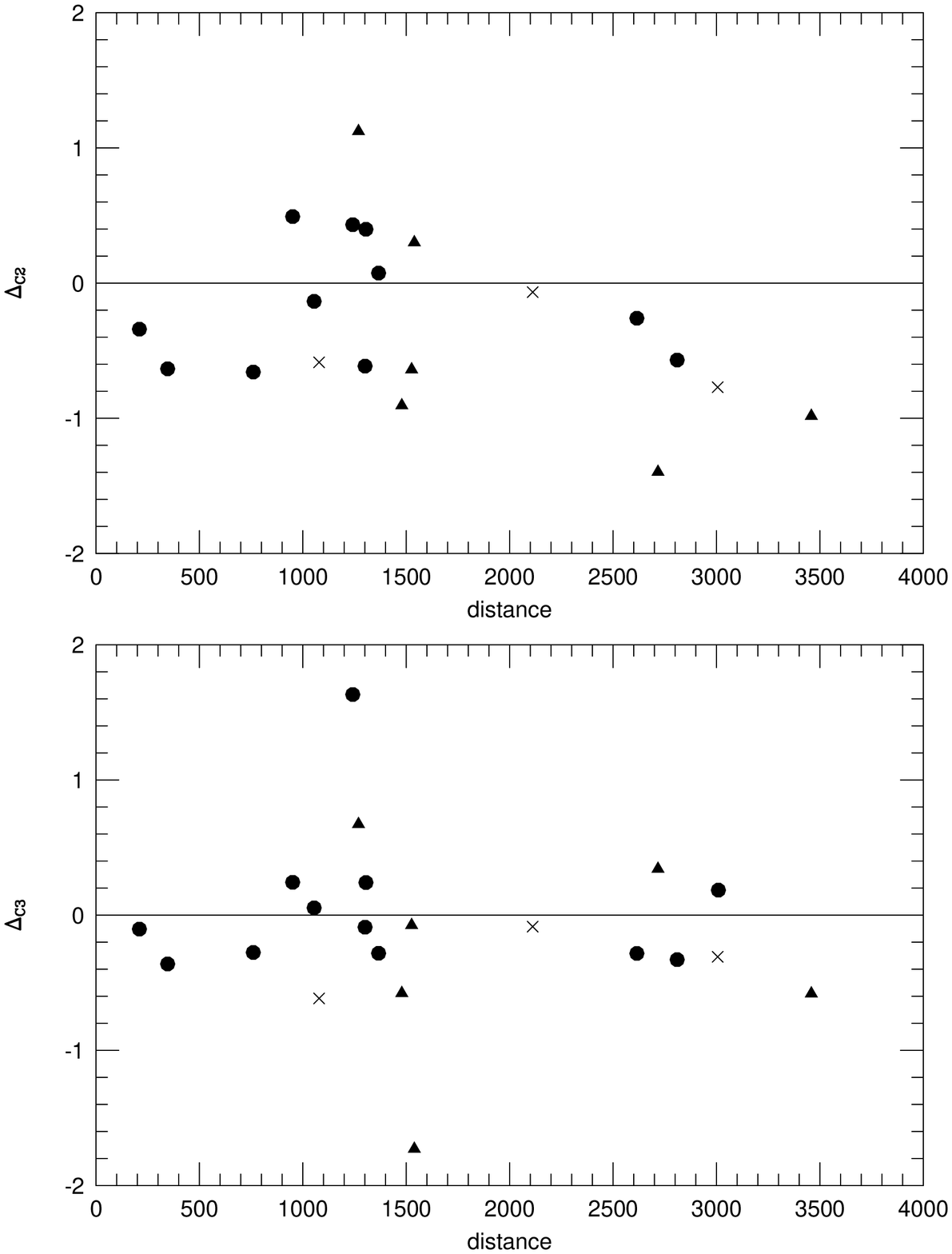}{7truein}{0}{80}{80}{-250}{-50}
\caption{
The normalised CaH indices, $\Delta_{C2}$ and $\Delta_{C3}$ 
plotted against distance above (or below) the Galactic Plane.
\label{fig6}
}
\end{figure}


\clearpage
\begin{thebibliography}{DUM}

\bibitem[Allard \& Hauschildt 1995] {ah95} Allard, F., Hauschildt, P.H. 1995, \apj, 445, 433
\bibitem[Bessell 1982] {b82}  Bessell, M.S. 1982, Proc. ASA, 4, 417
\bibitem[Cohen {\sl et al} 1996] {c96} Cohen, J.G., Hogg, D.W., Pahre, M.A., Blandford, R. 1996,
 \apj, 462, L9
\bibitem[Cohen {\sl et al} 1997] {c97} Cohen, J.G. et al 1997, in preparation
\bibitem[Cowie {\sl et al} 1996] {co96} Cowie, L.L., Songaila, A., Hu, E.M., Cohen, J.G. 1996, \aj, 
112, 839
\bibitem[Gilmore \& Reid, 1983] {gr83} Gilmore, G., Reid, N. 1983, \mnras, 202, 1025
\bibitem[Gilmore {\sl et al} 1995] {gwj95} Gilmore, G.F., Wyse, R.F.G., Jones, J.B. 1995, \aj, 109, 1095
\bibitem[Gizis 1997] {g97} Gizis, J.E. 1997, \aj, in press
\bibitem[Landolt 1992] {l92} Landolt, A.U. 1992, \aj, 104, 340
\bibitem[Larsen \& Humpherys 1996] {lh96} Larsen, J.A., Humphreys, R.M. 1996, \apj, 468, L99
\bibitem[Laureijs {\sl et al} 1994] {la94} Laureijs, R.J., Helou, G., Clark, F.O. 1994, in
First Symposium on the Infrared Cirrus and Diffuse Interstellar Clouds, ASP Conf. Ser. 58, 133,
ed. R. M. Cutri \& W. B. Latter
\bibitem[Majewski 1992] {m92} Majewski, S.R. 1992, \apjs, 78, 87
\bibitem[Majewski 1993] {m93}  Majewski, S.R. 1993, \araa, 31, 575
\bibitem[Norris 1987] {n87} Norris, J. 1987, \aj, 93, 616
\bibitem[Oke 1990] {o90} Oke, J.B. 1990, \aj, 99, 1621
\bibitem[Ojha {\sl et al} 1996] {o96} Ojha, D.K., Bienaym\'e, O., Robin, A.C., Cr\'ez\'e, M., Mohan, V. 1996,
\aap, 311, 456
\bibitem[Pahre {\sl et al} 1997] {p97} Pahre, M.A. et al 1997, in preparation
\bibitem[Reid \& Majewski 1993] {rm93} Reid, N., Majewski, S.R. 1993, \apj, 409, 635 (RM93)
\bibitem[Reid {\sl et al} 1995] {rhj95}  Reid, I.N., Hawley, S.L., Gizis, J.E.  1995, \aj , 110, 1838 (RHG)
\bibitem[Reid {\sl et al} 1996] {r96} Reid, I.N., Yan, L., Majewski, S.R., Thompson, I., Smail, I. 1996, 
\aj, 112, 1472 (RYMTS)
\bibitem[Robin {\sl et al} 1992] {r92} Robin, A.C., Cr\'ez\'e, M., Mohan, V. 1992, \aap, 265, 32
\bibitem[Robin 1994] {r94} Robin, A.C. 1994, IAU Symposium 161, Astronomy from Wide-field Imaging,
p. 403, ed. H.T. MacGillivray et al, Kluwer Academic Publishers, Dordrecht
\bibitem[Robin {\sl et al} 1996] {ro96} Robin, A.C., Haywood, M., Cr\'ez\'e, M., Ojha, D.K., Bienaym\'e, O.
1996, \aap, 305, 125
\bibitem[Trefzger {\sl et al} 1995] {t95} Trefzger, Ch.F., Pel, J.W., Gabi, S. 1995, \aap, 304, 381
\bibitem[Wyse \& Gilmore 1988] {wg88} Wyse, R.F.G., Gilmore, G.F. 1988, \aj, 95, 1404
\bibitem[Yoss {\sl et al} 1987] {y87} Yoss, K.M., Neese, C.L., Hartkopf, W.I. 1987, \aj, 94, 1600
\end{thebibliography}
\end{document}